\documentclass[letter]{aa}  

\usepackage{graphicx}
\usepackage{txfonts}
\usepackage{xcolor}
\usepackage{xspace}
\usepackage{hyperref}
\usepackage{comment}
\usepackage{multirow}
\hypersetup{
     colorlinks   = true,
     citecolor    = blue
}

\def \inte {{\em INTEGRAL\xspace}}
\def \ibis  {{\em IBIS/ISGRI\xspace}}
\def \swift {{\em Swift\xspace}}
\def \chandra {{\em Chandra\xspace}}

\def \igr{{IGR\,J17591$-$2342\xspace}}

\def \swiftxrt{{\em Swift/XRT\xspace}}
\def \nustar{{\em NuSTAR\xspace}}
\def \nicer{{\em NICER\xspace}}

\def \atca{{\em ATCA}}

\begin{document}

\title{\nustar{} and \nicer{} reveal \igr{} as a new accreting millisecond X-ray pulsar}

   \author{A. Sanna\inst{1},
          C. Ferrigno\inst{2},
          P. S. Ray\inst{3},
          L. Ducci\inst{4},
          G. K. Jaisawal\inst{5},
          T. Enoto\inst{6,7},
          E. Bozzo\inst{2},
          D. Altamirano\inst{8},
          T. Di Salvo\inst{9},
          T. E. Strohmayer\inst{10},
          A. Papitto\inst{11},
          A. Riggio\inst{1},
          L. Burderi\inst{1},
          P. M. Bult\inst{10},
          S. Bogdanov\inst{12},
          A. F. Gambino\inst{9},
          A. Marino\inst{13,14},
          R. Iaria\inst{9},
          Z. Arzoumanian\inst{10},
          D. Chakrabarty\inst{15},
          K. C. Gendreau\inst{10},
          S. Guillot\inst{16,17},
          C. Markwardt\inst{10},
          M. T. Wolff\inst{3} 
          }

   \institute{
   Dipartimento di Fisica, Universit\`a degli Studi di Cagliari, SP Monserrato-Sestu km 0.7, 09042 Monserrato, Italy\\
   		\email{andrea.sanna@dsf.unica.it}
   	    \and
	   ISDC, Department of Astronomy, University of Geneva, Chemin d'\'Ecogia 16, CH-1290 Versoix, Switzerland
        \and
        Space Science Division, Naval Research Laboratory, Washington, DC 20375-5352, USA
         \and
         Institut f\"ur Astronomie und Astrophysik, Eberhard Karls Universit\"at, Sand 1, 72076 T\"ubingen, Germany
        \and
        National Space Institute, Technical University of Denmark, Elektrovej 327-328, DK-2800 Lyngby, Denmark
         \and
         Department of Astronomy, Kyoto University, Kitashirakawa-Oiwake-cho, Sakyo-ku, Kyoto, Kyoto 606-8502, Japan
        \and
        The Hakubi Center for Advanced Research, Kyoto University, Yoshida-Ushinomiya-cho, Sakyo-ku, Kyoto, Kyoto 606-8302, Japan
        \and
        Physics \& Astronomy, University of Southampton, Southampton, Hampshire SO17 1BJ, UK
        \and
        Universit\`a degli Studi di Palermo, Dipartimento di Fisica e Chimica, via Archirafi 36, 90123 Palermo, Italy
        \and
        Astrophysics Science Division, NASA's Goddard Space Flight Center, Greenbelt, MD 20771, USA
        \and
        INAF, Osservatorio Astronomico di Roma, Via di Frascati 33, I-00044, Monteporzio Catone (Roma), Italy
        \and
        Columbia Astrophysics Laboratory, Columbia University, 550 West 120th Street, New York, NY 10027, USA
        \and
        INAF/IASF Palermo, via Ugo La Malfa 153, I-90146 - Palermo, Italy
        \and
        IRAP, Universit\`e de Toulouse, CNRS, UPS, CNES, Toulouse, France
        \and
        MIT Kavli Institute for Astrophysics and Space Research, Massachusetts Institute of Technology, Cambridge, MA 02139, USA
        \and
        CNRS, IRAP, 9 avenue du Colonel Roche, BP 44346, F-31028 Toulouse Cedex 4, France
        \and
        Universit\'e de Toulouse, CNES, UPS-OMP, F-31028 Toulouse, France
             }

   \date{Received -; accepted }

  \abstract
   {We report on the discovery by the {\em Nuclear Spectroscopic Telescope Array} (\nustar{}) and 
   the Neutron Star Interior Composition Explorer (\nicer{}) of the accreting millisecond X-ray pulsar \igr{}, detecting coherent X-ray pulsations around 527.4\,Hz (1.9\,ms) with a clear Doppler modulation. This implies an orbital period of $\sim 8.8$ hours and a projected semi-major axis of $\sim1.23$ lt-s. From the binary mass function, we estimate a minimum companion mass of $0.42$ M$_{\odot}$, obtained assuming a neutron star mass of 1.4 M$_\odot$ and an inclination angle lower than 60 degrees, as suggested by the absence of eclipses or dips in the light-curve of the source. 
   The broad-band energy spectrum is dominated by Comptonisation of soft thermal seed photons with a temperature of $\sim 0.7$\,keV by electrons heated to 21\,keV. We also detect black-body-like thermal direct emission compatible with an emission region of a few kilometers and temperature compatible with the seed source of Comptonisation. A weak Gaussian line centered on the iron K$\alpha$ complex can be interpreted as a signature of disc reflection. A similar spectrum characterises the \nicer{} spectra, measured during the outburst fading.
   }


\titlerunning{\igr{}: a new AMXP in outburst}
\authorrunning{Sanna et al.}

   \maketitle

\section{Introduction}
\label{sec:introduction}
Accreting, rapidly-rotating neutron stars (NS) in low mass X-ray binaries have been extensively investigated for almost two decades. This class of objects, also known as accreting millisecond X-ray pulsars (AMXPs), includes at the moment 21 sources with spin periods ranging between 1.7\,ms and 9.5\,ms \citep[see][for extensive reviews]{Burderi13,Patruno12b,Campana2018a}. The characteristic short spin periods observed in AMXPs are the result of long-lasting mass transfer from an evolved sub-Solar companion star via Roche-lobe overflow onto a slow-rotating NS \citep[{\it recycling scenario};][]{Alpar82}, making them the progenitors of rotation-powered millisecond pulsars shining from the radio to the gamma-ray band.
Almost a third of the AMXPs are \emph{ultra-compact} binary systems (P$_\mathrm{orb} < 1$ hrs), while the rest show on average short orbital periods (P$_\mathrm{orb} < 12$ hrs), except for the intermittent pulsar Aql\,X$-$1 \citep{Casella08} with $P_\mathrm{orb}\sim18$ hrs \citep{Welsh2000a}. Short orbital periods suggest small low-mass companion stars, consistent with donor masses on average <0.2~M$_{\odot}$.

Here, we report on the detection of millisecond X-ray pulsations from \igr{}, an X-ray transient first detected by the INTernational Gamma-Ray Astrophysics Laboratory (\inte{}) during a Galactic Center scanning on 2018 August 10 \citep{Ducci2018aa}. An archival search in the Neil Gehrels Swift Observatory (\swift{}) Burst Alert Telescope data revealed the source to be active since 2018 July 22 \citep{Krimm2018aa}. Pointed \swiftxrt{} observations of the source region, on 2018 August 12, revealed a point-like X-ray source 
and determined its precise X-ray position \citep{Bozzo2018aa},
which was refined by a \chandra{} observation on 2018 August 23 to be 
R.A.=$17^h 59^m 02.83^s$ Decl.=$-23^\circ 43' 10.2\arcsec$ (J2000)
with an astrometric uncertainty of 0.6\arcsec{} \citep{Nowak2018aa}. On 2018 August 14, the Australia Telescope Compact Array (\atca{}) detected \igr{} with a flux density $S_\nu = 1.09\pm0.02$ $\mu$Jy  and $S_\nu = 1.14\pm0.02$ $\mu$Jy (at 68\% c.l.) at 5.5 GHz and 9.0 GHz, respectively, at a position R.A. = $17^h 59^m 02.86^s \pm 0.04^s$ Decl.=$-23^{\circ} 43' 08.3\arcsec \pm 0.1\arcsec$, consistent with the X-ray determinations \citep{Russell2018aa}. 

In this letter, we describe a coherent timing analysis of the \nustar{} and \nicer{} observations that provided the pulsar spin period and binary ephemeris. We also report on the analysis of the X-ray spectral modelling obtained from \nustar{}, \swift{}, \inte{}, and \nicer{} data.

\section[]{Observations and data reduction}


\igr{} was observed by IBIS/ISGRI 
onboard the INTEGRAL satellite 
from 2018 August 10 at 15:50 to August 11 at 12:30 UT, for a total exposure time of 66\,ks.
We performed reduction and data analysis using the 
off-line science analysis (OSA) software
provided by the \inte{} Science Data Centre.
We produced a mosaic image from the combination of the
individual images of the available data set, where the source 
was detected with a significance  $>$7$\sigma$
in the 20$-$80\,keV energy band. We extracted the average IBIS/ISGRI spectrum in 8 bins from 25 to 150 keV with equal 
logarithmic spacing.
No JEM-X data were available due to the
source location at the edge of its field of view.

\nustar{} observed \igr{} (Obs.ID. 90401331002) on 2018 August 13 starting from 22:36 UT for a total exposure of $\sim30$ ks. We performed standard screening and filtering of the events using the \nustar{} data analysis software (\textsc{nustardas}) from \textsc{Heasoft} version 6.24. We extracted source events from a circular region of radius 80\arcsec{} centered on the source position. Due to straylight we extracted the background with a similar extraction area but located in a region far from the source with similar degree of stray-light contamination.
The average source count rate per instrument is $\sim9$ counts/s in the energy range 3$-$80 keV. We did not detect any Type-I 
thermonuclear burst during the observation.
We corrected for spacecraft clock drift applying the up-to-date clock correction file (version 84, valid up to 2018-08-14).

\swift{} observed \igr{} (Obs.ID. 00010804002) on 2018 August 14 from 00:38 UT for a total of $\sim0.6$ ks with \swiftxrt{} operated in photon counting mode. We reduced and processed the XRT data with \textsc{xrtpipeline} version 0.13.4, extracting source events from a circular region of radius 64\arcsec. We estimated a source count rate of $\sim{1}$ cts/s above the pile-up threshold \citep[see, e.g.][]{Romano2006aa}. 
Thus, we extracted the source spectrum using an annular region centered at the source position with inner and outer radius of 10\arcsec{} and 64\arcsec, respectively. Similarly, we extracted the background spectrum from an annular region with inner and outer radius of 102\arcsec{} and 230\arcsec, respectively.

\nicer{} observed \igr{} on 2018 August 15 from 00:00 to 14:08 UT for a total exposure of $\sim 7.3$\,ks (Obs.ID. 1200310101, hereafter \nicer{}-1), on August 18 from 02:08 to 03:56 UT (exposure of $\sim1.5$\,ks, Obs.ID. 1200310102, \nicer{}-2), on August 23 from 00:58 to 22:57 UT (exposure of $\sim11.5$\,ks, Obs.ID. 1200310103, \nicer{}-3) and on August 24 from 01:51 to 14:44 UT (exposure of $\sim7$\,ks, Obs.ID. 1200310104, \nicer{}-4). We filtered events in the 1--12\,keV band applying standard screening criteria with \textsc{NICERDAS} version 4.0. We removed short intervals showing background flaring in high geomagnetic latitude regions. The spectral background was obtained from blank-sky exposures. We did not detect any Type-I thermonuclear burst during the observations.
\section{Results}
\subsection{Timing analysis}
\label{sec:timing}


\begin{figure}
\centering
\includegraphics[width=0.45\textwidth]{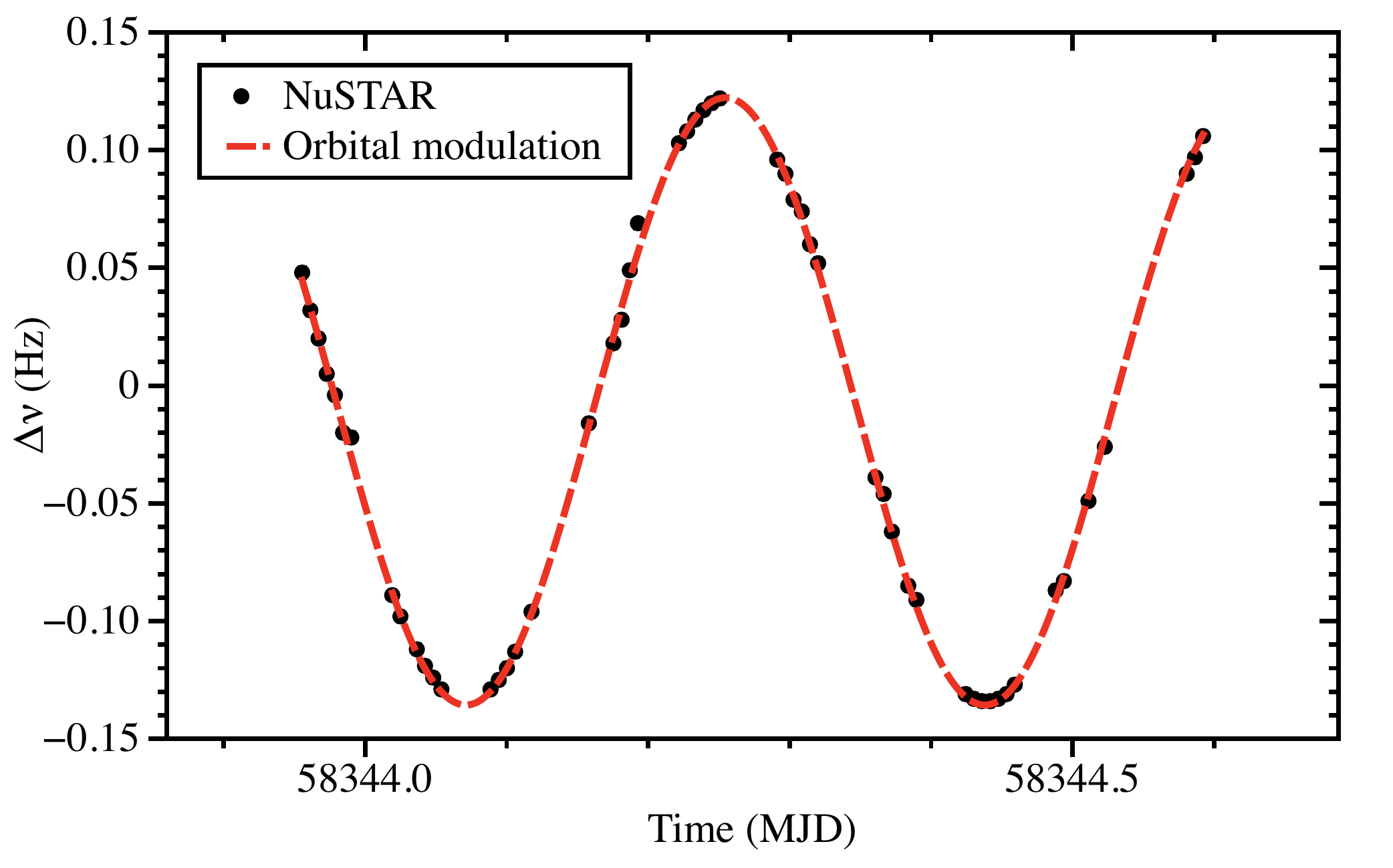}
\caption{Time evolution of the pulsar frequency shifts (with respect to $\nu=527.432$ Hz) estimated from 500-s \nustar{} data segments. The frequency shifts show a clear modulation compatible with Doppler shifts due to orbital motion in a binary system. The red dashed line represents the best-fitting orbital Doppler modulation model assuming a circular orbit.}
\label{fig:doppler}
\end{figure} 
To search for coherent signals, we first corrected the \nicer{} and \nustar{} photon arrival times to the Solar System barycenter with the tool \textsc{barycorr}, using the DE-405 Solar planetary ephemeris, and adopting the source coordinates of Tab.~\ref{tab:solution}. Power density spectra (PDS) obtained averaging either \nustar{} or \nicer{} 150-s data segments 
showed a prominent (>4$\sigma$) excess at a frequency of 527.3 Hz. The feature exhibited a double-peaked structure characteristic of orbital Doppler broadening. Inspection of the \nustar{} PDS over short segments (500-s) revealed pulse frequency modulation typical for pulsars in binary systems (see Fig.~\ref{fig:doppler}). The best-fit is found for orbital period $P_\mathrm{orb}=31694(67)$ seconds, projected semi-major axis of the NS orbit $x = 1.233(9)$ lt-s, time of passage through the ascending node $T_{NOD}=58343.704(1)$ (MJD) and spin frequency $\nu_0=527.4253(6)$ Hz.   

Starting from this provisional solution, we folded data segments of $\sim$400-s and $\sim$300-s for the \nustar{} and \nicer{} observations, respectively, into 8 phase bins at the preliminary spin frequency $\nu_0$.  
%
%
We modelled each pulse profile with a constant plus two sinusoidal functions, representing the pulse frequency and its second harmonic. We retained only profiles for which the ratio between the pulse frequency amplitude and its 1~$\sigma$ uncertainty was larger than three. Separately for \nustar{} and \nicer{}, we modelled the pulse phase with a linearly varying frequency plus a circular Keplerian orbital model. A more detailed description of the procedure can be found in \citet[][and references therein]{Sanna2016a}. 

\begin{figure}
\centering
\includegraphics[width=0.45\textwidth]{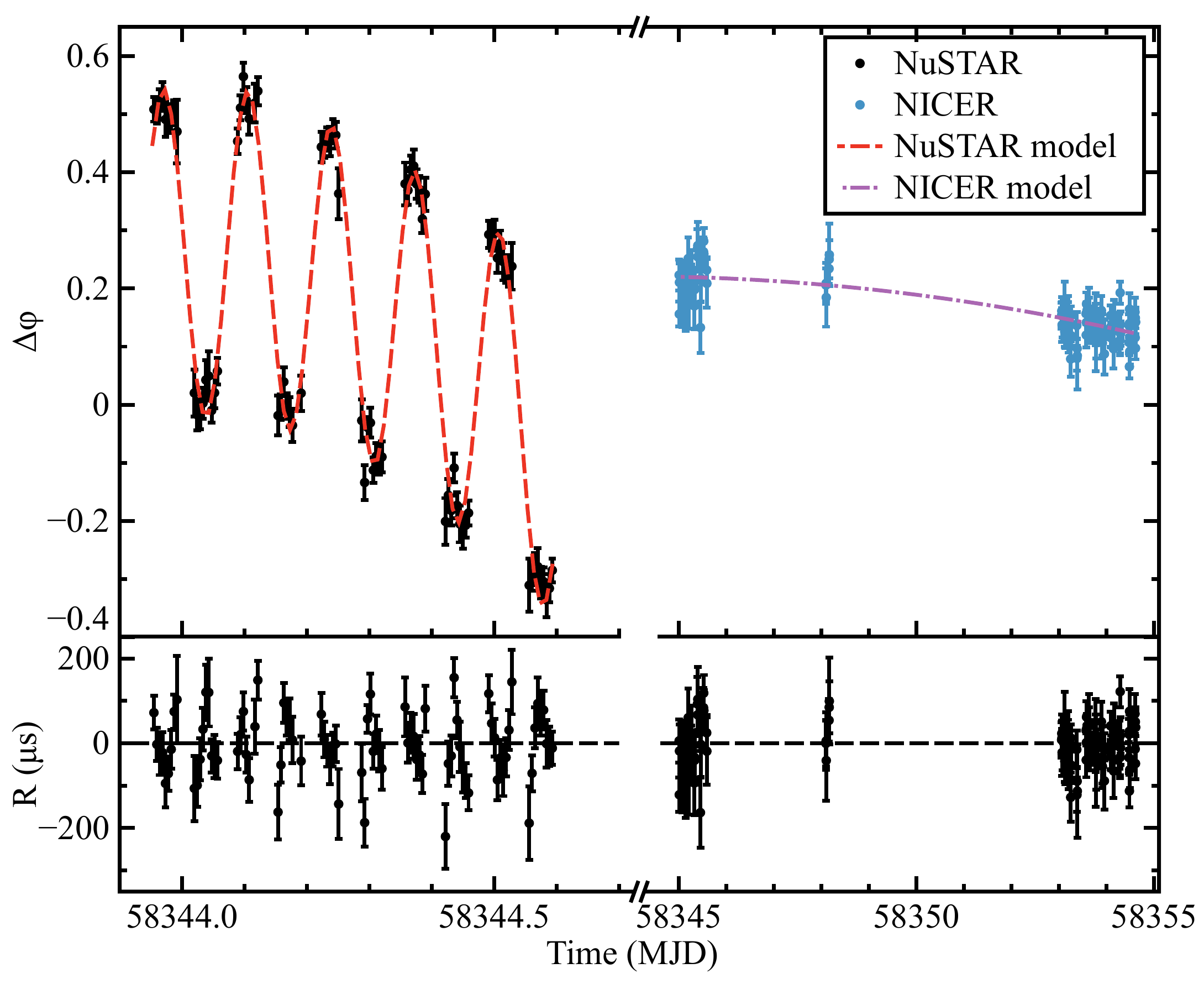}
\caption{\textit{Top panel -} Pulse phase delays as a function of time computed by epoch-folding at a constant frequency $\sim$400-s and 300-s data intervals of the \nustar{} (left) and the \nicer{} (right) data-sets, respectively, after correcting the best orbital solution. In the same panel we show the best-fit models for each group of data. \textit{Bottom panel -} Residuals in $\mu$s with respect to the best-fitting models for the pulse phase delays.}
\label{fig:phase_fit}
\end{figure}

The best-fit orbital and pulsar spin parameters for the \nustar{} and \nicer{} data-sets are shown in the first and second column of Tab.~\ref{tab:solution}, respectively. These are derived from the phase delays at the fundamental frequency; timing analysis using the second harmonic yields compatible results. The top panel of Fig.~\ref{fig:phase_fit} shows the \nustar{} and \nicer{} pulse phases after correcting for the best orbital solution and assuming a constant frequency. The \nustar{} pulse phases show a large distribution of residuals not compatible with the binary orbital modulation. These can be well described (see residuals in the bottom-panel left-side of Fig.~\ref{fig:phase_fit}) by a strong spin frequency derivative (see Tab.~\ref{tab:solution}) and an additional sinusoidal term with period $P\sim193$ minutes (close to twice the \nustar{} orbital period). We note that no sinusoidal modulation is present in either \nicer{} data (both harmonics) or the \nustar{} second harmonic pulse phases, suggesting a likely instrumental origin. For the latter, we emphasise that the phase variations of $\sim$0.5 observed for the fundamental component are close to a complete cycle of the harmonic, tuning down any effect of the phase drift. Note also that the lower statistics in the harmonic would make residuals less evident with respect to the fundamental. 

The \nicer{} pulse phase residuals (top-panel right side) show a hint of positive spin frequency derivative
(see Tab.~\ref{tab:solution} for the corresponding value and the bottom-panel right side of Fig.~\ref{fig:phase_fit} for final residuals accounting for this component).

\begin{figure}
  \includegraphics[width=0.48\textwidth]{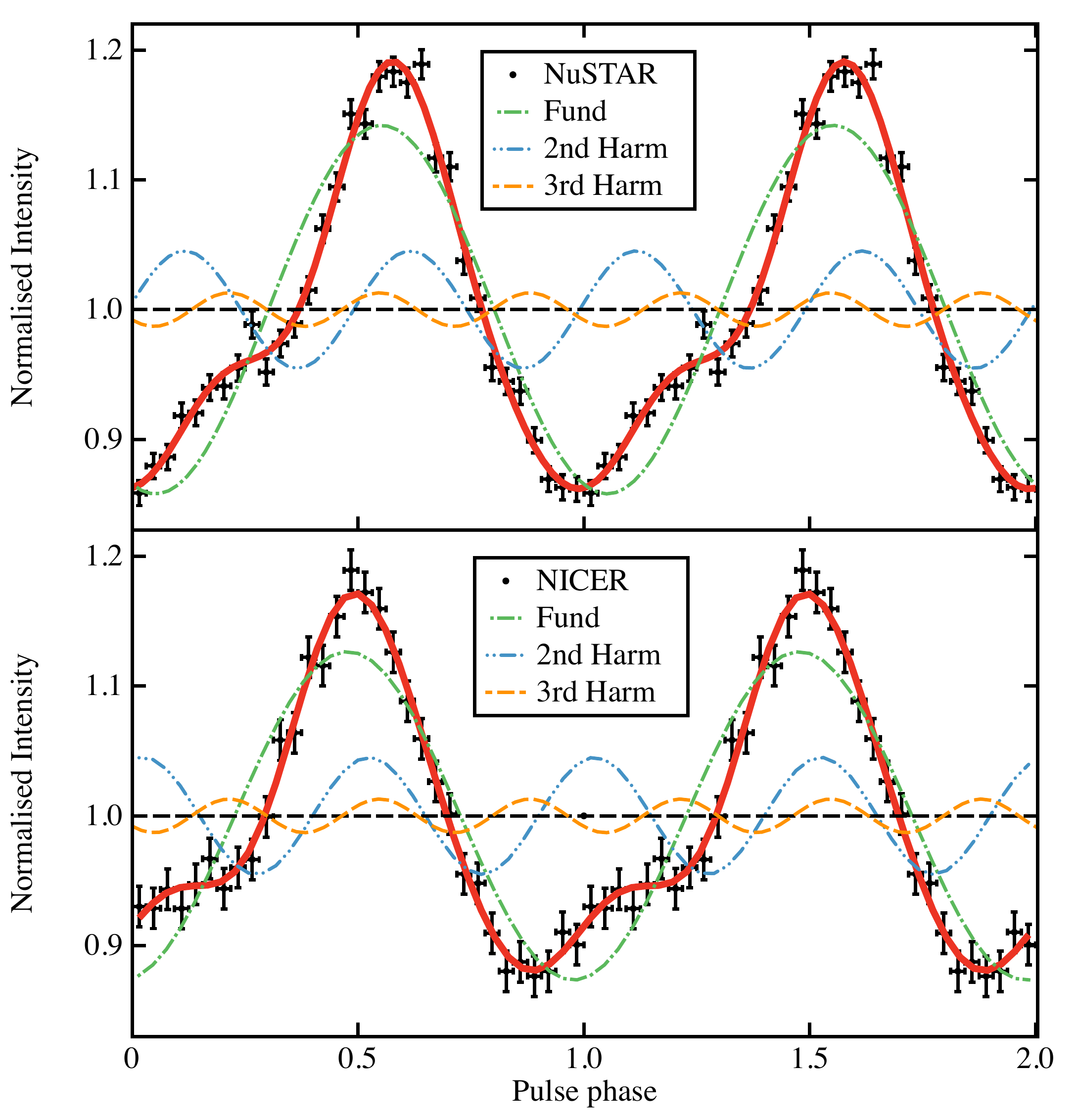}
  \caption{\igr{} pulse profiles (black points) from epoch-folding the \nustar{} (top-panel) and the \nicer{} (bottom-panel) observations after correcting for the orbital parameters reported in Tab.~\ref{tab:solution}. The best-fitting model (red line) is the superposition of three sinus functions with harmonically related periods. For clarity, we show two cycles of the pulse profile.}     
  \label{fig:profile}
  \end{figure}

We improved the orbital parameter estimates (third column of Tab.~\ref{tab:solution}) by orbitally phase-connecting the available data-sets \citep[see, e.g.][for more details on the procedure]{Sanna2017b, Sanna2017ab}. We note that the uncertainty on the source localisation has an effect on the determination of the frequency and frequency derivative which is 
 more than one order of magnitude smaller than the uncertainties reported in Tab.~\ref{tab:solution} for the corresponding parameters. Therefore, we ignore this effect.
Finally, we created pulse profiles by epoch-folding the \nustar{} (3--80 keV) and \nicer{} (1--12 keV) observations with the parameters reported in Tab.~\ref{tab:solution}. We modelled them with a combination of three sinusoidal components (see Fig.~\ref{fig:profile}). In \nustar{} data, the fundamental, the second and the third harmonics have fractional amplitudes of $(16.1\pm0.3)$\% , $(5.1\pm0.3)$\% and $(1.5\pm0.3)$\%, respectively. \nicer{} pulse profiles are characterised by fractional amplitudes of $(13.7\pm0.2)$\%, $(5.0\pm 0.2)$\% and $(1.0\pm0.2)$\% for fundamental, the second and the third harmonics, respectively. 

\begin{table}
\caption{Orbital parameters and spin frequency of \igr{} with uncertainties on the last digit quoted at 1$\sigma$ confidence level.}
\centering
\resizebox{0.49\textwidth}{!}{
\begin{tabular}{l c c c }
Parameters  & \nustar{} & \nicer{} & \nustar{}+\nicer{} \\
\hline
\hline
R.A.\tablefootmark{a} (J2000) &  \multicolumn{3}{c}{$17^h59^m02.86^s \pm 0.04^s$}\\
Decl.\tablefootmark{a} (J2000) & \multicolumn{3}{c}{$-23^\circ43' 08.3\arcsec \pm 0.1\arcsec$}\\
$P_\mathrm{orb}$ (s) &31684.8$\pm$0.1&31684.743$\pm$0.003&31684.738$\pm$ 0.002\\
$x$ (lt-s) &1.22774$\pm$1$\times 10^{-5}$&1.227716$\pm$8$\times 10^{-6}$&1.227728$\pm$7$\times 10^{-6}$\\
$T_{NOD}$ (MJD) & 58345.171984$\pm$4$\times 10^{-6}$&58345.1719787$\pm$16$\times 10^{-7}$&58345.1719786$\pm$14$\times 10^{-7}$\\
e &$ < 1 \times 10^{-4}$ & $< 6 \times 10^{-5}$&$< 5.5 \times 10^{-5}$\\
$\nu_0$ (Hz) &527.425790$\pm$1$\times 10^{-6}$&527.42570042$\pm$8$\times 10^{-8}$&--\\
$\dot{\nu}$ (Hz/s) & \tablefootmark{b}(2.6$\pm$0.3)$\times 10^{-10}$ &(2.0$\pm$1.6)$\times 10^{-13}$&--\\
$T_0$ (MJD) & 58344.0 & 58344.0& 58344.0\\
\hline
$\chi^2_\mathrm{red}$/d.o.f. & 2.27/65 &1.25/99 & 1.67/177\\
\end{tabular}
}
\tablefoot{
\tablefoottext{a}{Localisation of the ATCA radio counterpart \citep{Russell2018aa}.} \tablefoottext{b}{The spin frequency derivative likely originates from the time drift of the internal clock of the \nustar{} instrument, see Sec.~\ref{sec:discussion} for more details.} $T_0$ represents the reference epoch for these timing solutions. Frequencies and times reported here are referred to the TDB time scale. }

\label{tab:solution}
\end{table}

\subsection{Spectral analysis}
\label{sec:spectra}
\begin{figure}
 \resizebox{0.49\textwidth}{!}{
  \includegraphics[angle=0]{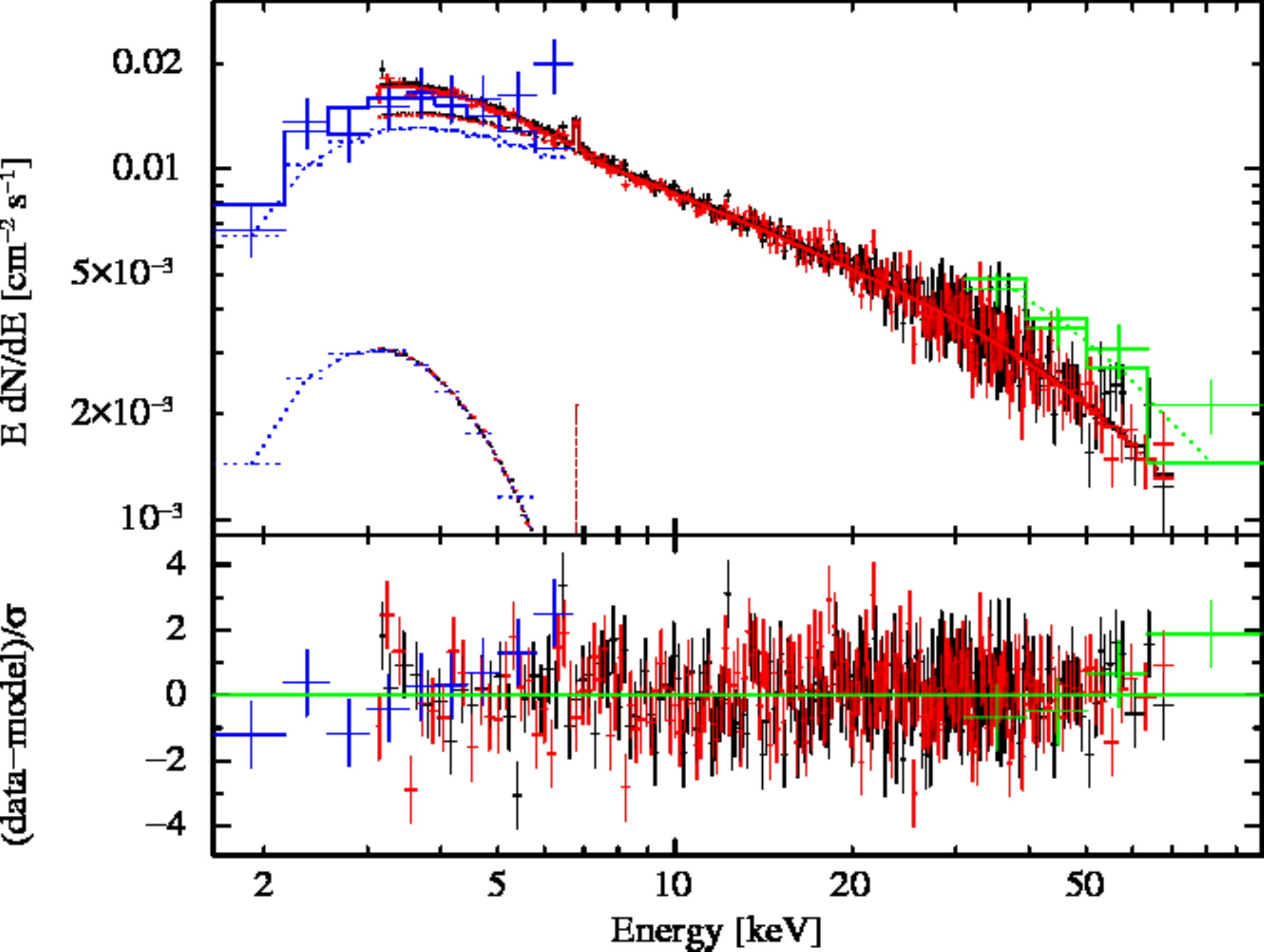}
  }
  \caption{\emph{Upper panel}: \swift{} (blue), \nustar{} FPMA and FPMB (black and red, respectively) and \ibis{} (green) unfolded spectra of \igr{}. The dashed and dash-dotted lines represent black-body and Comptonisation components respectively. The sharp dotted line is a Gaussian with width set to zero. Additional rebinning is for displaying purposes. \emph{Lower panel}: residuals from the best-fit model in units of standard deviations.}
  \label{fig:spectrum}
\end{figure}

We performed spectral analysis with Xspec 12.10.0c \citep{Arnaud96} after applying an optimal binning \citep{Kaastra2016aa}. For the first epoch, we used the 0.5--7.5\,keV range for \swiftxrt{}, 3--70 keV for \nustar{} and 30--80\,keV for \ibis{} and grouped the spectra to collect at least 20 counts per bin (Fig.~\ref{fig:spectrum}). 
The broad-band (0.5--80 keV) spectrum is well-described by an absorbed thermal component (black-body) with temperature $kT_\mathrm{BB}$ that provides also the seed photons for a thermally Comptonised continuum with electron energy $kT_\mathrm{e}$ and a weak emission line centered on the iron K$\alpha$ complex and width fixed to zero. In Xspec, this is implemented as 
\texttt{TBabs*[cflux*nthcomp+bbodyrad]+gauss}.
For modelling, we left the flux of the Comptonisation component free to vary independently for each instrument, to account for cross-instrument calibration offsets and the possible source variations due to not strictly simultaneous observations.

The \nicer{} observations were fitted independently by minimising the \texttt{pgstat} statistic in Xspec\footnote{\texttt{pgstat} is suited for Poisson distributed data with Gaussian distributed background; see \url{https://heasarc.gsfc.nasa.gov/xanadu/xspec/manual/XSappendixStatistics.html}.}. Due to uncertainties in the low-energy response, we ignored the \nicer{} data below 1.4\,keV. 
We adopted the same model, but we fixed the electron temperature at the value found from \nustar{} and checked that the result remained insensitive as long as $kT_\mathrm{e}>12$\,keV.
After verifying the compatibility of single spectra, we performed a joint fit of \nicer{}-1 and 2 observations, leaving only the flux to be independently determined between them. We applied the same procedure to \nicer{}-3 and 4. 
The best-fit parameters are shown in Tab.~\ref{tab:spectral_fit} with uncertainties at the 90\% confidence level. 
Regardless of the statistical test used to find the best-fit parameters, we reported the $\chi^2_\mathrm{red}$ value.


\begin{table}\label{tab:spectral_fit}
\caption{Best-fit spectral parameters of \igr{} with uncertainties at 90\% confidence level.}
    \centering
    \resizebox{0.45\textwidth}{!}{
    \begin{tabular}{l r@{}l r@{}l r@{}l}
    \hline
    \hline
    Parameter & \multicolumn{2}{c}{multiple\tablefootmark{a}} & \multicolumn{2}{c}{\nicer{}-1,2\tablefootmark{a}} & \multicolumn{2}{c}{\nicer{}-3,4\tablefootmark{a}} \\
\hline
\smallskip
$N_\mathrm{H}$ [$10^{22}$cm$^{-2}$] & 3.6 & $\pm$1.1 &	3.45 & $^{+0.18}_{-0.15}$ &	3.59 & $^{+0.05}_{-0.08}$ \\ 
\smallskip
$\Gamma$ & 1.76 & $\pm$0.02 &	1.83 & $\pm$0.06 &\ 	2.23 & $\pm$0.05 \\ 
\smallskip
$kT_\mathrm{e}$ [keV] & 22 & $^{+4}_{-3}$ &	 22&(fixed) &	22&(fixed) \\
\smallskip
$kT_\mathrm{BB}$ [keV] & 0.79 & $\pm$0.09 &	 0.58 & $^{+0.05}_{-0.07}$ & 0.90 & $^{+0.07}_{-0.06}$ \\ 
\smallskip
$r_\mathrm{bb}$ [(km / D$_\mathrm{10kpc}$)$^2$] &2.6 & $^{+1.4}_{-0.6}$ &	 3.3 & $^{+0.8}_{-0.6}$   &	 $<5\times$& $10^{3}$ \\ 
\smallskip
$E_\mathrm{Fe}$ [keV] & 6.82 & $^{+0.14}_{-0.39}$ &	 6.35 & $\pm0.04$  &  6.37 & $\pm$0.10 \\
\smallskip
$N_\mathrm{Fe}$ [$10^{-5}$ph/s/cm$^{-2}$] & 5.2 & $\pm1.8$ & 3.8& $\pm$0.9 &	 2.9 & $\pm$1.6 \\
\smallskip
\multirow{2}{*}{Flux (1--10\,keV)\tablefootmark{b}} & \multirow{2}{*}{\tablefootmark{c}2.0} & \multirow{2}{*}{$\pm0.3$} & 	\tablefootmark{d}1.85 & $\pm$0.09 	&  \tablefootmark{e}2.35 & $^{+0.03}_{-0.04}$ \\ \smallskip
 & & & 	\tablefootmark{d}1.06 & $^{+0.08}_{-0.09}$ 	&  \tablefootmark{e}2.42 & $^{+0.02}_{-0.04}$ \\ 
\smallskip
Flux (3--20\,keV)\tablefootmark{b} & 2.41 & $\pm0.05$ & -&- &	 -&- \\
\smallskip
Flux (20--100\,keV)\tablefootmark{b} & 3.8 & $^{+0.4}_{-0.5}$ & 	 -&- &	 -&- \\
\smallskip
\multirow{2}{*}{Flux (0.1--10\,keV)\tablefootmark{f}} & \multirow{2}{*}{1.73} & 	  & 1.31\tablefootmark{d}&	&1.57\tablefootmark{e} & \\
 &  & & 0.78\tablefootmark{d} &  &	1.61\tablefootmark{e} & \\
$\chi^2_\mathrm{red}$/d.o.f. & 1.09 &/473 &	 1.11 &/377 &\ 	 1.13 &/1129 \\ 
\hline
    \end{tabular}
    }
    \tablefoot{
\tablefoottext{a}{``multiple'' indicated the combined fit of the \nustar, \ibis, and \swiftxrt{} observations. \nicer{}-1 and \nicer{}-1,2 are the observations of 2018, August 15 and 18. \nicer-3,4 of August 23 and 24.}
\tablefoottext{b}{The flux of the \texttt{nthComp} component in units of $10^{-10}\,\mathrm{erg\,s^{-1}\,cm^{-2}}$. The 3--20 and 20--100\,keV \texttt{nthcomp} 
fluxes apply to \nustar{} and \ibis{} data, respectively. \ibis{} flux is higher that the others, since the data were obtained earlier, when the source was brighter.}
\tablefoottext{c}{This flux applies to \swiftxrt{} data only.}
\tablefoottext{d}{The upper and lower rows refer to the \nicer{}-1 and 2 observations.}
\tablefoottext{e}{The upper and lower rows refer to the \nicer{}-3 and 4 observations.}
\tablefoottext{f}{Extrapolated absorbed flux derived by the model in units of $10^{-10}\,\mathrm{erg\,s^{-1}\,cm^{-2}}$. For the multiple observation, we extrapolate the \nustar{} spectrum.}
}
\end{table}

\section{Discussion}
\label{sec:discussion}
We reported on the newly discovered AMXP \igr{}, for which we detected coherent X-ray pulsations at $\sim527$~Hz in the \nustar{} and \nicer{} observations performed almost 25 days from the beginning of the outburst, with a pulse fraction of 15\%. We modelled the NS spin frequency drift as the Doppler shift induced by the binary orbital motion, discovering the binary nature of the system, characterised by an orbital period of almost 8.8 hours, very similar to the intermittent AMXP SAX J1748.9$-$2021 \citep[see, e.g.][]{Altamirano2008a, Sanna2016a} and the eclipsing AMXP SWIFT J1749.4$-$2807 \citep[see, e.g.][]{Markwardt2010aa, Altamirano2011a, Ferrigno2011a}.
\noindent

The analysis of the \nustar{} pulse phase residuals revealed the presence of a large spin-up derivative ($2.6\times 10^{-10}$ Hz/s) and an oscillation close to twice the satellite orbital period. We suggest that both effects are associated with the time drift of the internal clock instrument \citep[see, e.g.][]{Madsen15}. Similar spurious frequency derivatives have been reported for \nustar{} observations of AMXPs such as MAXI\,J0911$-$655 \citep{Sanna2017a}, SAX J1808.4$-$3658 \citep{Sanna2017ab} and IGR\,J00291+5934 \citep{Sanna2017b}. Also the large discrepancy on the spin frequency (($\sim1\times 10^{-4}$ Hz) between  \nustar{} and \nicer{} is a direct consequence of the \nustar{} clock issue. These considerations are reinforced by the analysis of \nicer{} observations, which does not evidence any significant frequency drift.


A spin-up frequency derivative of (2.0$\pm$1.6)$\times 10^{-13}$\,Hz/s is observed from the phase-coherent timing analysis of the \nicer{} observations.
For a broad-band (0.1--100 keV) absorbed flux of $\sim7\times10^{-10}$ erg/s/cm$^2$ and a source distance of 8.5 kpc \citep[assumed near the Galactic Center, see, e.g.][]{Kerr1986aa}, we estimate an accretion rate of $\dot{M}\simeq5.2\times10^{-10}$ M$_{\odot}$/yr (for a NS radius and mass of 1.4\,M$_{\odot}$ and 10 km, respectively). Assuming the accretion disc to be truncated at the co-rotation radius, the observed mass accretion rate yields a maximum spin-up derivative of a few $10^{-13}$ Hz/s, fully consistent with the observed one. 

 

A rough estimate of the NS dipolar magnetic field can be obtained by assuming the condition of spin equilibrium for the X-ray pulsar. The magnetic field can then be estimated as:
\begin{equation}
\label{eq:spineq}
B=0.63\,\zeta^{-7/6}\left(\frac{P_{\text{spin}}}{2\text{ms}}\right)^{7/6}\left(\frac{M}{1.4M_{\odot}}\right)^{1/3}\left(\frac{\dot{M}}{10^{-10}M_{\odot}/\text{yr}}\right)^{1/2}10^8 \text{G},
\end{equation}
where $\zeta$ represents a model-dependent dimensionless factor (between 0.1--1) corresponding to the ratio between the magnetospheric radius and the Alfv\'en radius \citep[see, e.g.,][]{Ghosh79a,Wang96}, $P_{\text{spin}}$ is the pulsar spin period in ms and $M$ is the NS mass.
Assuming a 1.4 M$_\odot$ NS and the value of $\dot{M}$ reported above, we obtain a range for the dipolar magnetic field of $1.4\times10^8<B<8\times 10^{9}$ G, consistent with the average magnetic field of known AMXPs \citep[see, e.g.][]{Mukherjee2015}.

\begin{figure}
  \includegraphics[width=0.46\textwidth]{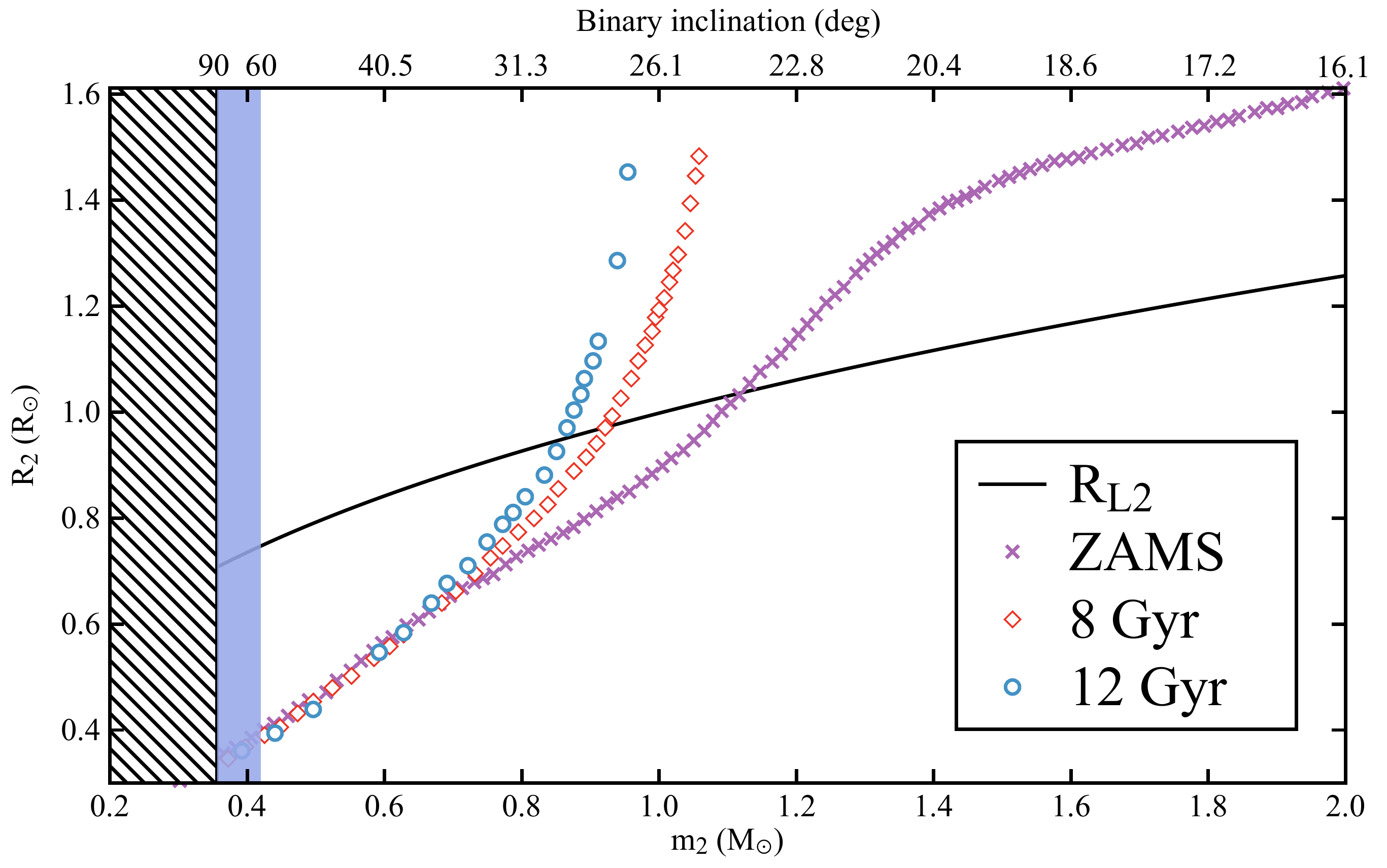}
  \caption{Radius-mass plane showing the size constraints on the companion star Roche-Lobe of \igr{} (black solid line) obtained from the orbital parameters of the neutron star. The hatched region represents the constraints on the companion mass from the binary mass function, while the blue area defines the mass constraints for inclination angles between 60 and 90 degrees. The other curves represent theoretical mass-radius relations for Zero-Age Main Sequence stars (purple crosses) and isochrones for stars with age of 8 (red diamonds) and 12\,Gyr (blu circles). The top x-axes represents the corresponding binary inclination angle in degrees assuming a 1.4 M$\odot$. }     
  \label{fig:mass}
\end{figure}

From the NS mass function $f(m_2, m_1, i)\sim1.5 \times 10^{-2}$~M$_{\odot}$, we can constrain the mass of the companion star. Since neither total eclipses, nor dips, have been observed in the light curve, we can assume a binary inclination lower than 60 degrees 
\citep[see, e.g.][]{Frank02}. As shown in Fig.~\ref{fig:mass}, the upper limit on the inclination angle (represented in blue) allows us to limit the companion star mass $m_2 \gtrsim $0.42~M$_{\odot}$ (for a 1.4~M$_{\odot}$ NS), which increases up to $m_2 \gtrsim 0.52$~M$_{\odot}$ if we consider a 2~M$_{\odot}$ NS. Introducing the contact condition ($R_2\approx R_{L2}$) required to activate Roche-Lobe overflow, we can express the donor radius as a function of its mass as $R_2\simeq 0.87\,m_2^{1/3}\,P_\mathrm{orb,9h}^{2/3}$ R$_{\odot}$, where $m_2$ is the companion mass in units of M$_{\odot}$ and $P_\mathrm{orb,9h}$ represents the binary orbital period in units of 9 hours. In Fig.~\ref{fig:mass}, we report the companion mass-radius relation (black solid line) assuming a 1.4 M$_{\odot}$ NS. For comparison, we show numerical simulated mass-radius relations for Zero-Age Main Sequence stars (ZAMS) \citep[purple crosses; see, e.g.][]{Tout1996aa}, as well as isochrones for stars aged 8 (red diamonds) and 12\,Gyr \citep[blue circles; see, e.g.][]{Girardi2000aa}. From the intersections with the mass-radius companion curve, we can infer that the donor is compatible with either a ZAMS with mass $\sim1.1$ M$_\odot$ (corresponding to an inclination angle of $i\sim24$ degrees) or an old main sequence star with mass 0.85$-$0.92 M$_\odot$ ($i$ ranging between $28$ and $30$ degrees) for a star age between 8 and 12\,Gyr. We note, however, that the \textit{a priori} probability of observing a binary system with inclination $i\leq 30$ degrees is of the order of 13\%. Mass values for which the main sequence radius is smaller than the companion Roche-lobe could still be acceptable if we consider the possibility of a bloated donor star. In that case, the thermal timescale ($GM^2_2/R_2 L_2$) should be much larger than the evolutionary timescale ($M_2/\dot{M_2}$).


Finally, the broad-band energy spectrum of \igr{} is well described by an absorbed soft black-body like component ($kT\sim 0.8$ keV) with relatively small emitting area compatible with emission from the neutron star surface (or part of it) plus a Comptonised component ($\Gamma \sim 1.8$) with a seed photon temperature compatible with the soft thermal component. The source spectral properties are consistent with other AMXPs observed in the hard state \citep[see, e.g.][]{Falanga05a,Gierlinski2005a, Papitto09, Papitto2013a,Sanna2017a,Sanna2017b}. We found marginal evidence ($\sim4\sigma$, based on an F-test) of a weak emission line compatible with the iron K-$\alpha$ transition. Even if marginally significant, its introduction removes
positive residuals around the
expected energy and such lines are not unusual for this kind of sources  \citep[see, e.g.][]{Papitto09,Papitto2013a, Sanna2017a, Sanna2017b}. From Tab.~\ref{tab:spectral_fit}, we note that at later times, when the source has a re-brightening, the additional black body seems to disappear, while the asymptotic Comptonisation power-law index increases. This might be due to more effective cooling of the seed photons by a 
thicker accretion stream above the stellar surface.


The discovery of another transient by \inte{} and its characterisation as the 22$^\mathrm{nd}$ accreting millisecond pulsar by \nustar{} and \nicer{} enriches the census of these key objects 
in the understanding of the late stages of stellar evolution.

\begin{acknowledgements}
We thank Fiona Harrison for accepting a \nustar{} observation in the Director Discretionary Time. We thank the \swift{},\nustar{}, and \nicer{} teams for scheduling and performing these target of opportunity observations on a short notice. This work was supported by NASA through the \nicer{} and \nustar{} missions, and the Astrophysics Explorers Program. It made use of data and software provided by the High Energy Astrophysics Science Archive Research Center (HEASARC). It is also based on observations with \inte{}, an ESA project with instruments and science data centre funded by ESA member states (especially the PI countries: Denmark, France, Germany, Italy, Switzerland, Spain) and with the participation of Russia and the USA. 
DA acknowledges support from the Royal Society. LD acknowledges grant FKZ 50 OG 1602.
\end{acknowledgements}

\bibliographystyle{aa} 
\bibliography{biblio.bib}

\end{document}